\documentclass[10pt]{article}
\usepackage{amsmath}
\usepackage{textcomp}
\usepackage{pstricks}
\usepackage{pst-node}
\usepackage{prooftree}
\usepackage{fullpage}
\usepackage{coqdoc}
\usepackage{url}

\newcommand{\Coq}{\textsc{Coq}}
\newcommand{\leut}{\ensuremath{=<\,}}
\newcommand{\lfcv}{\ensuremath{\scriptsize <<\!\!-\,}}
\newcommand{\rfcv}{\ensuremath{\scriptsize-\!\!>>}}
\newcommand{\conva}{\mbox{\lfcv\coqdocid{a}\rfcv}}
\newcommand{\iconva}{\ensuremath{<--\!\!} \coqdocid{a} \ensuremath{--\!\!>}}

\title{(Mechanical) Reasoning\\on Infinite Extensive  Games\\[12pt]\textbf{\normalsize LIP report RR2008-16}}

\author{Pierre Lescanne\\
Universit\'e de Lyon, ENS de Lyon, CNRS (LIP), \\
46 all\'ee d'Italie, 69364 Lyon, France}
\begin{document}

\maketitle{}

\pagestyle{empty}
\thispagestyle{empty}

\begin{abstract}
  In order to better understand reasoning involved in analyzing infinite games in extensive form, we performed the experiments in proof assistant \Coq{} that are reported here.
\end{abstract}

\section{Introduction}









One of the main aims of game theory is to understand how agents reason.  Psychologists~\cite{stahl95:_player_model,colman03:_depth_of_strat,zhang03:_two_parad} would say agents are
human and tries to answer the question of \emph{how} human agents reason. In this paper, we take a radically different view; for us, agents are ideal abstract entities with
unlimited formal deduction power and we attempt to answer the question of \emph{what} a full reasoning can be.  For that we decided to analyze the process in full detail on a
mechanical device, namely a proof assistant run on a computer which performs at the extreme level of detail all the steps of reasoning.  This way, we hope to be able to highlight
concepts and deductions that are necessary and that a human would do more or less.  Among the possible concepts involved in a reasoning, we claim that some are forgotten whereas
irrelevant others are considered.  In the first class is temporal reasoning (``always'', ``eventually''), in the second is the use of ``excluded middle'' or ``double
negation''.  Our experiments have shown that far from easy deductions are used\footnote{The reader is invited to have a look at the scripts.}, in particular with infinite games.
This complexity may explain why human reasoning departs from what is expected.

Among the parts of game theory that have been overlooked, is this of infinite games, on which formal reasoning is rather subtle.  In this paper we report research about the concepts
underlying infinite games and experiments on the proof assistant \Coq~\cite{barras00:_coq_proof_assis_refer_manual} to make formal reasoning effective.

In this paper, we are presenting mechanical reasoning that is deductions that, unlike human ones, are based on the fundamental principles of logic.  This means that every piece of
reasoning has to be justified by a rule of logic and therefore the whole process can be checked by a machine, i.e., a computer, the \Coq{} software in our case.  The domain of
experience is this of \emph{games in extensive form} (\cite{osborne04a} Chap.~5), that are sequential games were each situation are owned by a player, whose turn comes one after
the others.  More specifically, we focus on infinite such games, where there may be infinitely many situations.

\section{COQ and the Constructive Logic}
\label{sec:coq_const_logic}

Around 1980 a new concept called \textit{Curry-Howard correspondence} emerged. It relies on type theory and lambda-calculus~\cite{GirardLafontTaylor89} and says basically that
\textit{proofs are programs}.  In this theory, all objects have a type, that is an annotation that limits its use.  For instance, an object $f$ of type $A\rightarrow B$, written $f:A\rightarrow B$
represents a function and can only be applied to an object of type $A$ to produce an object of type $B$.  In our development we will write, for instance, $a:Agent$ to say that $a$
is an \emph{Agent} and $s:InfStrategy$ to say that $s$ is an infinite strategy.  In what follows, a node (written \textit{iNode}) in a infinite strategy is something that takes an
agent, a choice, a finite strategy and an infinite strategy and produces an infinite strategy has a type, in more precise words it has the type,
\[\coqdocid{iNode}: Agent \rightarrow  Choice \rightarrow  FinStrategy \rightarrow  InfStrategy \rightarrow  InfStrategy.\]   
The Curry-Howard correspondence says also that \emph{types are propositions} and insists essentially on the computational content of proofs and establishes the bases of the so-called constructive logic.  Indeed since a proof
works as a computation, an object can be taken into consideration only if it can be constructed and an existential proof is accepted if it allows constructing the object it
claims the existence of.  For instance, in an infinite extensive game we assert that there exists a utility for an agent $a$ that can be associated with an infinite strategy, but this
works if we prove it exists, i.e., provide a way to construct this utility, since assuming the existence without the construction is not enough.

The following formal development in \Coq{} has been built in this framework and checked on a computer and is attached to proof scripts available on the web site of the
author\footnote{\url{http://perso.ens-lyon.fr/pierre.lescanne/COQ/INFGAMES/}}.  In this article we try to describe and comment the content of the scripts without entering in their
full technicality, but the reader is anyway strongly encouraged to have a look at the scripts to convince himself of the materiality of the proofs and moreover, if he has access to
a \Coq{} implementation, he should try to run them on a computer.


\subsection{Natural deduction}
\label{sec:natural}
The proof systems we describe is based on natural deduction, which is one of the main system to formalize logic.  Natural deduction has been
formalized by Gerhard Gentzen~\cite{gentzen35} in 1935 (see also~\cite{praw:natu65,Dalen97}).  Its adjective ``natural'' comes form the fact that its
creators considered that it is just the natural way to formalize logic.  It is based on the fact that to conduct a proof, one works under
hypotheses and one tries to draw a conclusion.  One considers that a theorem has been proved when all the hypotheses have been discharged.

The basic concept of natural deduction is this of \emph{sequent}.  A sequent is a pair of a context $\Gamma$ and a proposition $\varphi$, which is written $\Gamma\vdash\varphi$ and which means that
$\varphi$ is a logical consequence of the set of hypotheses $\Gamma$.  A context is a set of propositions and when we write $\Gamma, \varphi$ we mean that the sequent $\Gamma$ is enriched by the
proposition $\varphi$, in other terms this is a notation for the union of sets of propositions.  In this calculus there is one axiom and several rules.  The axiom is
\begin{displaymath}
  \Gamma \vdash\varphi \qquad \textrm{if~} \varphi\in \Gamma
\end{displaymath}
There are two kinds of rules for each connector, namely \emph{introduction rule} and \emph{elimination rule}.  For instance, for the implication
$\rightarrow $, the introduction rule is
\begin{displaymath}
  \prooftree
  \Gamma, \varphi \vdash \psi
\justifies \Gamma\vdash \varphi\rightarrow  \psi
\endprooftree
\end{displaymath}
whereas the elimination rule is 
\begin{displaymath}
  \prooftree
\Gamma\vdash \varphi\rightarrow  \psi \qquad \qquad \Gamma\vdash \varphi
\justifies \Gamma\vdash \psi
  \endprooftree
\end{displaymath}
also known as the \emph{modus ponens}.
We are now giving only the rule for the universal quantifier, the introduction rule 
\begin{displaymath}
  \prooftree
\Gamma, x:A \vdash P(x)
\justifies \Gamma \vdash \forall x:A, P(x)
  \endprooftree %
\end{displaymath}
and the elimination rule:
\begin{displaymath}
  \prooftree
\Gamma \vdash a:A \qquad \Gamma \vdash \forall x:A, P(x)
\justifies \Gamma \vdash  P(a) 
\endprooftree
\end{displaymath}

Through the Curry-Howard correspondence, natural deduction is strongly connected to type theory and computation.

\subsection{COQ}
\label{sec:coq}

In this paper we are not describing all the details of the proof assistant \Coq{} and the book~\cite{BertotCasterant04} is excellent source of information.  Basically in \Coq{}, proofs
are mathematical objects in natural deduction and are considered as first class citizen that can be printed, be exchanged as objects between people, and overall be checked by a
proof checker (a specific software) that examines it in full detail.  As building proofs is a tedious process, the software \Coq{} offers tools to build them.  Basically a \Coq{}
development is made of several not fully separated phases: the user defines data structures (in our case: games and strategies), then he defines predicates, relations and functions
on those structures (in our case: conversion, utilities and equilibria) and finally he proves theorems about those structures and predicates (in our case: a theorem that says that
subgame perfect equilibria are Nash equilibria). Typically a script presents a set of sections in which a subdevelopment is presented, this section may invoke other sections; it
is a sequence of declaration of variables, axioms, or hypothesis, followed by
definitions and theorems with their proofs.  To check that a script is correct, it is highly recommended to run it on a \Coq{} implementation.

\section{Induction, Coinduction, and  Fixed point}
\label{sec:fixpt}

\emph{Induction} is a tool to reason over infinite sets of objects, provided those objects are finitely based.  On the opposite, \emph{coinduction}~\cite{Bertot_induction} has been
designed to reason on infinite objects, like games with infinite paths.  However infinite objects may have parts that are finite like finite paths in infinite games and induction
can also be used on infinite objects, specifically on their finite subparts.  In particular, if the paths are finite, one can compute the utility.  Moreover in infinite games we can
define finite relations, called \textit{convertibility}.  Therefore finite concepts are interleaved with infinite ones.

Beside games, we can define other finite or infinite objects, further we consider the following \emph{inductive} (finite) objects: 1) finite games, 2) finite strategies (or strategy
profiles), 3) the predicate \emph{eventually right} that says when applied to a path of a game that the path in question goes eventually to the right, 4) the two convertibility
congruences among strategies, namely among finite strategies, but also among infinitely strategy (we indeed claim this latter congruence can be described finitely provided we
restrict it to appropriate strategies). 5) Nash equilibrium on finite games, 6) backward induction predicate and 7) Nash equilibrium on infinite games are also finite inductive concepts.

In this paper the archetype of a \emph{coinductive} object is an infinite game.  Obviously the associated concept of infinite strategy is coinductive.  The function
\coqdocid{i2u} that associates a utility with a strategy is also coinductive since the infinite strategies are.  The predicate \coqdocid{SGPE} which tells whether an infinite
strategy is a \emph{subgame perfect equilibrium} is a coinductive object since the strategy is. In what follows we will define predicates that say that a property is ``always'' satisfied
along a path, those properties on infinite objects are coinductively defined.

Roughly speaking inductive definitions are like equations and correspond to define the concept as \emph{least fixed points} and the properties of this concept are derived of this
minimality. On another hand, coinductive definitions correspond to \emph{greatest fixed points} and the properties of this concept are derived from the maximality.  Handling infinite objects (actual infinity vs potential infinity) is tricky and the \Coq{} user does not escape this rule. Notice that \Coq{}
offers tools to verify along a proof that the one one builds will be accepted by the checker.

\section{Informal presentation of infinite games}
\label{sec:informal}


In this paper, we analyze \emph{games in extensive form}.  Informally such games are presented as trees.  Each node of the tree is a situation in the game where a
player has to take a decision.  For reason of simplicity, following Vestergaard~\cite{vestergaard06:IPL}, we studied \emph{binary games} where the players have only two choices.
This seems a reasonable design decision, since we can reduce a choice
among $n$ to a binary choice between the $n^{th}$ and the set $\{1,..., n-1\}$ followed by the binary choice between the $(n-1)^{th}$ and the set $\{1,...,n-2\}$ etc...  We do not loose any
generality in term of modeling, and in term of abstraction this is the same as $n$ choices at each node.  A reader interested by an implementation based on more than two choices, i.e., on polyadic trees, is advised to consult St\'{e}phane Le~Roux PhD~\cite{LeRouxPhD08}.  By considering binary games, we hope to be more didactic.

First we will consider \emph{finite binary games}.  Those games are somewhat connected with games with \emph{finite horizon}, but in addition to offer choices to a finite depth,
they are also finitely branching.  Such a game is either a leaf, that is a game which ends and attributes the utilities, or a node where an agent has to take a decision.  This
leads typically to an inductive description.  Like a natural is either $0$ or the successor of a natural, or a binary tree is either a leaf with a content or a node made of two
binary subtrees, a finite game is either a leaf with a function that associates a utility to each agent or a node with an agent and two subgames, namely a left subgame and a right
subgame.  With a finite game, we associate strategies. A~strategy has the same structure as the game it is associated with, except that to each (internal) node we give a direction
\emph{left} or \emph{right}, which corresponds to the choice made by the agent in this particular situation.  In other words, a strategy is either a leaf associated with a utility,
exactly like for games, or a node with an agent, a choice (left or right), and two substrategies.

The main goal of the study presented in this paper is infinite games, more specifically we describe deductions and formal reasoning on those games.  Again for reason of
simplicity, we study infinite games that looks like ``centipedes'', in other words binary games with a unique infinite path.  The basic concept of those games is a node made of
three components: an agent, a left subgames which is itself an infinite subgame, and a right subgame which is a finite subgame.  This means that we ``recycle'' the formal development
made for finite subgames for the right branch.  Since the game is infinite its formalization relies no more on induction, but on coinduction.  Like for finite binary games, the
attached concept of strategy relies on a node with four components: an agent, a choice (left or right), an infinite left substrategy, and a finite right strategy.

\section{Equilibria on finite Games}
\label{sec:eq_fin_ga}

As an introduction to the \Coq{} development, let us study equilibria on \emph{finite games}.

\subsection{Agent and Utility}
\label{sec:utilities}

First we set two basic concepts namely \coqdocid{Agent} and \coqdocid{Utility}. In \Coq, this is done by the following declaration:

\medskip
\noindent
\coqdockw{Variable} \coqdocid{Agent} : \coqdocid{Set}.\coqdoceol
\noindent
\coqdockw{Variable} \coqdocid{Utility}: \coqdocid{Set}.\coqdoceol 

\medskip 

In addition we set a \coqdocid{preference} as a binary relation on \coqdocid{Utility} and we make this \coqdocid{preference} a preorder, which we write \leut, \coqdocid{Utility\_fun} is the type of the \emph{utility functions}, in other words the type of the objects belonging to \coqdocid{Agent}
\ensuremath{\rightarrow} \coqdocid{Utility}.

\medskip

\subsection{Finite Games}
\label{sec:fin_games}

A finite binary game, which we call a \coqdocid{FinGame}, in \Coq{}, is built by induction, this means that this is either a \emph{leaf} or this is a game with two subgames that
are themselves binary games, we encapsulate these three items under the label \textit{gNode}.  In \Coq, such a  data structure will be written:

  \medskip
\noindent
\coqdockw{Inductive} \coqdocid{FinGame} : \coqdocid{Set} :=\coqdoceol
\noindent
| \coqdocid{gLeaf} : \coqdocid{Utility\_fun} \ensuremath{\rightarrow} \coqdocid{FinGame}\coqdoceol
\noindent
| \coqdocid{gNode} : \coqdocid{Agent} \ensuremath{\rightarrow} \coqdocid{FinGame} \ensuremath{\rightarrow} \coqdocid{FinGame} \ensuremath{\rightarrow} \coqdocid{FinGame}.\coqdoceol

\medskip\noindent where \coqdockw{Inductive} is the key word to introduce any inductive definition.  \coqdocid{FinGame} is a data structure, but elsewhere the same
\coqdockw{Inductive} keyword will be used to define a predicate.  An inductive definition creates a deduction rule that allows us to reason by induction, namely in the case of
\coqdocid{FinGame}, the term \coqdocid{FinGame\_ind} is created by \Coq:

\medskip

\ensuremath{\forall} \coqdocid{P} : \coqdocid{FinGame} \ensuremath{\rightarrow} \coqdocid{Prop},\coqdoceol
       (\ensuremath{\forall} u : \coqdocid{Utility\_fun}, \coqdocid{P} (\coqdocid{gLeaf} \coqdocid{u})) \ensuremath{\rightarrow}\coqdoceol
       (\ensuremath{\forall} (\coqdocid{a} : \coqdocid{Agent}) (\coqdocid{f0} :\coqdocid{FinGame}),
        \coqdocid{P} \coqdocid{f0} \ensuremath{\rightarrow} \ensuremath{\forall} \coqdocid{f1} : \coqdocid{FinGame}, \coqdocid{P} \coqdocid{f1} \ensuremath{\rightarrow} \coqdocid{P} (\coqdocid{gNode} a \coqdocid{f0} \coqdocid{f1})
              )\coqdoceol
            
          \ensuremath{\rightarrow}
       \ensuremath{\forall} \coqdocid{f1} : \coqdocid{FinGame}, \coqdocid{P} \coqdocid{f1}

\medskip
\noindent It can be written as the rule
\begin{displaymath}
  \prooftree
  \ensuremath{\forall} u : \coqdocid{Utility\_fun}, \coqdocid{P} (\coqdocid{gLeaf}\ \coqdocid{u}) %
  \qquad %
   [\ensuremath{\forall} \coqdocid{f0} :\coqdocid{FinGame},
  \coqdocid{P}\ \coqdocid{f0} \ensuremath{\wedge}  \ensuremath{\forall} \coqdocid{f1} : \coqdocid{FinGame}, \coqdocid{P}\ \coqdocid{f1}] \ensuremath{\rightarrow}  \ensuremath{\forall} \coqdocid{a} : \coqdocid{Agent}, \coqdocid{P}\ (\coqdocid{gNode}\ a\ \coqdocid{f0}\ \coqdocid{f1})
  \justifies \ensuremath{\forall} \coqdocid{f} : \coqdocid{FinGame}, \coqdocid{P}\ \coqdocid{f}
  \endprooftree
\end{displaymath}
\noindent in other words, to prove that some properties holds for all the finite binary games, one has to prove it to hold for leaves and to prove that if it holds for two games
then it holds for the game obtained by pairing those two games under the ``control'' of an agent.

\subsection{Finite Strategies}
\label{sec:fin_strat}

As we said, a finite strategy\footnote{Actually we should probably say a \emph{strategy profile}, but, for convenience and conciseness, we call this data structure just a \emph{strategy}.} has a
structure which is very similar to a finite game, the only difference is that we add a ``choice'' at each node.

\medskip
\noindent
\coqdockw{Inductive} \coqdocid{FinStrategy} : \coqdocid{Set} :=\coqdoceol
\noindent
| \coqdocid{sLeaf} : \coqdocid{Utility\_fun} \ensuremath{\rightarrow} \coqdocid{FinStrategy}\coqdoceol
\noindent
| \coqdocid{sNode} : \coqdocid{Agent} \ensuremath{\rightarrow} \coqdocid{Choice} \ensuremath{\rightarrow} \coqdocid{FinStrategy} \ensuremath{\rightarrow} \coqdocid{FinStrategy} \ensuremath{\rightarrow} \coqdocid{FinStrategy}.\coqdoceol
\medskip

With a strategy we provide a function that associates with every strategy a utility function.  The \Coq{} keyword to define such a function is \coqdockw{Fixpoint}.  A syntactic
construction called \coqdocid{match} allows using a pattern matching mechanism.

\medskip
\noindent
\coqdockw{Fixpoint} \coqdocid{f2u} (\coqdocid{s}:\coqdocid{FinStrategy}) : \coqdocid{Utility\_fun} :=\coqdoceol
\noindent
\coqdocid{match} \coqdocid{s} \coqdocid{with}\coqdoceol
\noindent
| (\coqdocid{sLeaf} \coqdocid{uf})            \ensuremath{\Rightarrow} \coqdocid{uf}\coqdoceol
\noindent
| (\coqdocid{sNode} \coqdocid{a} \coqdocid{left} \coqdocid{sl} \coqdocid{sr})  \ensuremath{\Rightarrow}  (\coqdocid{f2u} \coqdocid{sl})\coqdoceol
\noindent
| (\coqdocid{sNode} \coqdocid{a} \coqdocid{right} \coqdocid{sl} \coqdocid{sr}) \ensuremath{\Rightarrow}  (\coqdocid{f2u} \coqdocid{sr})\coqdoceol
\coqdocindent{1.00em}
\coqdocid{end}.\coqdoceol

This reads as
\begin{itemize}
\item if the strategy \coqdocid{s} is a leaf, i.e., \coqdocid{sLeaf} \coqdocid{uf} (where \coqdocid{uf} is a utility function), then one returns \coqdocid{uf},
\item otherwise one returns the utility function associated with the left substrategy, if the choice is \emph{left}, or the right substrategy if the choice is \coqdocid{right}.
\end{itemize}
Now we define on finite strategies a relation, which we call \coqdocid{a}-\emph{convertibility}, and which we write \coqdocid{s}\conva\coqdocid{s'}.  Labeled with
$a$, it is associated with the agent \coqdocid{a} and says that
\begin{itemize}
\item Two leaves associated with the same utility function are \coqdocid{a}-convertible,
\item Two \coqdocid{sNode}s, (\coqdocid{sNode} \coqdocid{a} \coqdocid{c} \coqdocid{s1} \coqdocid{s2}) and (\coqdocid{sNode} \coqdocid{a} \coqdocid{c'} \coqdocid{s1'}
  \coqdocid{s2'}), associated with the same agent \coqdocid{a} are \coqdocid{a}-convertible if \coqdocid{s1} \conva \coqdocid{s1'} and \coqdocid{s2} \conva
  \coqdocid{s2'}.  Notice that \coqdocid{c} and \coqdocid{c'} do not have to be the same,
\item Two \coqdocid{sNode}s, (\coqdocid{sNode} \coqdocid{a'} \coqdocid{c} \coqdocid{s1} \coqdocid{s2}) and (\coqdocid{sNode} \coqdocid{a'} \coqdocid{c} \coqdocid{s1'}
  \coqdocid{s2'}), associated with another agent \coqdocid{a'} are \coqdocid{a}-convertible if \coqdocid{s1} \conva \coqdocid{s1'} and \coqdocid{s2} \conva
  \coqdocid{s2'}.  Notice that in this case \coqdocid{c} has to be the same.
\end{itemize}
We prove in \Coq{} (a first interesting exercise) that the \coqdocid{a}-convertibility is an equivalence relation i.e., it is reflexive, symmetric and transitive.  We are now equipped to define
the predicate \emph{Nash equilibrium} on \emph{finite strategy} (in \Coq{} the predicate  \coqdocid{FinNashEq} on \coqdocid{FinStrategy}):

\medskip
\noindent
\coqdockw{Inductive} \coqdocid{FinNashEq}: \coqdocid{FinStrategy} \ensuremath{\rightarrow} \coqdocid{Prop} := \coqdoceol
\noindent
| \coqdocid{NE} : \ensuremath{\forall} (\coqdocid{s}:\coqdocid{FinStrategy}), \coqdoceol
\coqdocindent{1.00em}
(\ensuremath{\forall} (\coqdocid{a}:\coqdocid{Agent}) (\coqdocid{s'}:\coqdocid{FinStrategy}), \coqdocid{s}\conva\coqdocid{s'} \ensuremath{\rightarrow} (\coqdocid{f2u} \coqdocid{s'} \coqdocid{a} \leut \coqdocid{f2u} \coqdocid{s} \coqdocid{a})) \ensuremath{\rightarrow} \coqdoceol
\coqdocindent{1.00em}
\coqdocid{FinNashEq} \coqdocid{s}.\coqdoceol
\medskip

It says that \coqdocid{s} is a Nash equilibrium if for all strategy \coqdocid{s'} that is \coqdocid{a}-convertible to \coqdocid{s}, 
\begin{center}
  \coqdocid{f2u} \coqdocid{s'} \coqdocid{a} \leut \coqdocid{f2u} \coqdocid{s} \coqdocid{a},
\end{center}
in other words, the utility for \coqdocid{a} computed for \coqdocid{s'} is less than the utility for \coqdocid{a} computed for \coqdocid{s}.  This is nothing more than the
traditional definition of Nash equilibrium for extensive game, written  in the formalism of \Coq.

Beside the predicate \emph{Nash equilibrium}, we define the predicate \emph{BI} which says whether the strategy \coqdocid{s} can be obtained by the so-called \emph{backward induction}.
In \Coq, it is written:

\medskip

\noindent
\coqdockw{Inductive} \coqdocid{BI}: \coqdocid{FinStrategy} \ensuremath{\rightarrow} \coqdocid{Prop} :=\coqdoceol
\noindent
| \coqdocid{BILeaf}: \ensuremath{\forall} \coqdocid{uf}:\coqdocid{Utility\_fun}, \coqdocid{BI} (\coqdocid{sLeaf} \coqdocid{uf})\coqdoceol
\noindent
| \coqdocid{BINode\_left}: \ensuremath{\forall} (\coqdocid{a}:\coqdocid{Agent}) (\coqdocid{sl} \coqdocid{sr}: \coqdocid{FinStrategy}), \coqdoceol
\coqdocindent{2.00em}
\coqdocid{BI} \coqdocid{sl} \ensuremath{\rightarrow} \coqdocid{BI} \coqdocid{sr} \ensuremath{\rightarrow} (\coqdocid{f2u} \coqdocid{sr} \coqdocid{a} \leut \coqdocid{f2u} \coqdocid{sl} \coqdocid{a}) \ensuremath{\rightarrow}\coqdoceol
\coqdocindent{2.00em}
\coqdocid{BI} (\coqdocid{sNode} \coqdocid{a} \coqdocid{left} \coqdocid{sl} \coqdocid{sr})\coqdoceol
\noindent
| \coqdocid{BINode\_right}: \ensuremath{\forall} (\coqdocid{a}:\coqdocid{Agent}) (\coqdocid{sl} \coqdocid{sr}: \coqdocid{FinStrategy}), \coqdoceol
\coqdocindent{2.00em}
\coqdocid{BI} \coqdocid{sl} \ensuremath{\rightarrow} \coqdocid{BI} \coqdocid{sr} \ensuremath{\rightarrow} (\coqdocid{f2u} \coqdocid{sl} \coqdocid{a} \leut \coqdocid{f2u} \coqdocid{sr} \coqdocid{a}) \ensuremath{\rightarrow}\coqdoceol
\coqdocindent{2.00em}
\coqdocid{BI} (\coqdocid{sNode} \coqdocid{a} \coqdocid{right} \coqdocid{sl} \coqdocid{sr}).\coqdoceol
\medskip

In other words,
\begin{itemize}
\item  \emph{BI} holds for leaves,
\item if \emph{BI}  holds for \coqdocid{sl} and \coqdocid{sr} and \coqdocid{f2u} \coqdocid{sl} \coqdocid{a} \leut \coqdocid{f2u} \coqdocid{sr} \coqdocid{a} then \emph{BI}  holds for \coqdocid{sNode} \coqdocid{a} \coqdocid{left} \coqdocid{sl} \coqdocid{sr} (\coqdocid{BINode\_left} principle),
\item the \coqdocid{BINode\_right} principle is symmetric, it says, that if \emph{BI} holds for \coqdocid{sl} and for \coqdocid{sr} and \coqdocid{f2u}
  \coqdocid{sr} \coqdocid{a} \leut \coqdocid{f2u} \coqdocid{sl} \coqdocid{a} then is \emph{BI} holds for \coqdocid{sNode} \coqdocid{a} \coqdocid{right} \coqdocid{sl} \coqdocid{sr}.
\end{itemize}
This is the formalization of backward induction for finite horizon games as described in textbooks~\cite{osborne94:_cours_game_theory,osborne04a,gintis00:_game_theor_evolv}.  Then
we are able to prove in \Coq{} the theorem:

\medskip
\noindent
\coqdockw{Theorem} \coqdocid{BI\_is\_FinNashEq} : \ensuremath{\forall} \coqdocid{s}, \coqdocid{BI} \coqdocid{s} \ensuremath{\rightarrow} \coqdocid{FinNashEq} \coqdocid{s}.\coqdoceol

\medskip

The theorem relies on the following fact: the inductive definition of \emph{BI} is a somewhat equational definition that says that \emph{BI} is the least fixed point of that
equation. Therefore if we can prove that if \coqdocid{FinNashEq} is another fixed point, this other fixed point is implied by \emph{BI} and we are set.  This can be done by three
lemmas.

\medskip

$\bullet$ \coqdocid{FinNashEq} satisfies the statement given by  \coqdocid{BILeaf}:

\medskip
\noindent \coqdockw{Lemma} \coqdocid{FinNashEq Leaf} : \ensuremath{\forall} (\coqdocid{uf}:\coqdocid{Utility\_fun}), \coqdocid{FinNashEq} (\coqdocid{sLeaf} \coqdocid{uf}).\coqdoceol

\medskip

$\bullet$ \coqdocid{FinNashEq} satisfies the statement given by \coqdocid{BINode\_left}:

\medskip
\noindent \coqdockw{Lemma} \coqdocid{FinNashEq\_fixpt\_left} : \ensuremath{\forall} (\coqdocid{a}:\coqdocid{Agent}) (\coqdocid{s1} \coqdocid{s2}: \coqdocid{FinStrategy}),
\coqdoceol \coqdocindent{1.00em} \coqdocid{FinNashEq} \coqdocid{s1} \ensuremath{\rightarrow} \coqdocid{FinNashEq} \coqdocid{s2} \ensuremath{\rightarrow} (\coqdocid{f2u}
\coqdocid{s2} \coqdocid{a} \leut \coqdocid{f2u} \coqdocid{s1} \coqdocid{a}) \ensuremath{\rightarrow} \coqdoceol \coqdocindent{1.00em} \coqdocid{FinNashEq} (\coqdocid{sNode}
\coqdocid{a} \coqdocid{left} \coqdocid{s1} \coqdocid{s2}).\coqdoceol

\medskip

$\bullet$ \coqdocid{FinNashEq} satisfies the statement given by \coqdocid{BINode\_right}:

\medskip
\noindent \coqdockw{Lemma} \coqdocid{FinNashEq\_fixpt\_right} : \ensuremath{\forall} (\coqdocid{a}:\coqdocid{Agent}) (\coqdocid{s1} \coqdocid{s2}: \coqdocid{FinStrategy}),
\coqdoceol \coqdocindent{1.00em} \coqdocid{FinNashEq} \coqdocid{s1} \ensuremath{\rightarrow} \coqdocid{FinNashEq} \coqdocid{s2} \ensuremath{\rightarrow} (\coqdocid{f2u}
\coqdocid{s1} \coqdocid{a} \leut \coqdocid{f2u} \coqdocid{s2} \coqdocid{a}) \ensuremath{\rightarrow} \coqdoceol \coqdocindent{1.00em} \coqdocid{FinNashEq} (\coqdocid{sNode}
\coqdocid{a} \coqdocid{right} \coqdocid{s1} \coqdocid{s2}).\coqdoceol

\medskip

Then we conclude than \coqdocid{FinNashEq} satisfies the same equation as \coqdocid{BI}.  Since \coqdocid{BI} is the least fixed point, \coqdocid{BI} \coqdocid{s} implies
\coqdocid{FinNashEq} \coqdocid{s}.

\section{Coinduction}
\label{sec:coind}

Coinduction is the partner of induction, but whereas the induction defines objects or predicates as the least fixed point of some equations and therefore specifies finite objects
and presents them from basic objects, coinduction defines infinite objects or infinite \underline{and} finite objects as greatest fixed point of some equation.  If one has in mind
to define infinite objects only, there is no need to specify basic objects (i.e., objects meant to be the basis on which to define finite object).  The typical infinite object
is

\medskip
\noindent
\coqdockw{CoInductive} \coqdocid{InfGame} : \coqdocid{Set} :=\coqdoceol
\noindent
| \coqdocid{igNode} : \coqdocid{Agent} \ensuremath{\rightarrow} \coqdocid{InfGame} \ensuremath{\rightarrow} \coqdocid{FinGame} \ensuremath{\rightarrow} \coqdocid{InfGame}.\coqdoceol

\medskip

In other words, an infinite binary game is made of an agent, an infinite binary subgame, and a finite binary subgame.  On the same model as for finite binary games, we define infinite strategies the same way, i.e., as a coinductive:

\medskip

\noindent
\coqdockw{CoInductive} \coqdocid{InfStrategy} : \coqdocid{Set} :=\coqdoceol
\noindent
| \coqdocid{iNode} : \coqdocid{Agent} \ensuremath{\rightarrow} \coqdocid{Choice} \ensuremath{\rightarrow} \coqdocid{InfStrategy} \ensuremath{\rightarrow} \coqdocid{FinStrategy} \ensuremath{\rightarrow} \coqdocid{InfStrategy}.\coqdoceol

\section{Equilibria on Infinite Games}
\label{sec:eq_inf_ga}

\subsection{Decomposing Infinite Games}
\label{sec:dec_inf_games}

Since the strategies  are infinite we can define total functions \coqdocid{iAgent} (which gives the agent at the root of the game), \coqdocid{SubGameLeft} (the infinite game on the left) and \coqdocid{SubGameRight} (the finite game on the right) with the lemma:

\medskip
\noindent
\coqdockw{Lemma} \coqdocid{InfGame\_Decomposition}: \ensuremath{\forall} (\coqdocid{g}:\coqdocid{InfGame}),\coqdoceol
\coqdocindent{1.00em}
\coqdocid{igNode} (\coqdocid{iAgent}  \coqdocid{g}) (\coqdocid{SubGameLeft} \coqdocid{g}) (\coqdocid{SubGameRight} \coqdocid{g}) = \coqdocid{g}.\coqdoceol
\medskip 

\noindent that says that a game can be uniquely decomposed into an agent, a left subgame and a right subgame.  

\subsection{Generalizing the notion of horizon}
\label{sec:horizon} 

In classical extensive game theory, the concept of backward induction relies on this of \textit{finite horizon}, our experience in mechanizing the proof has shown us that finite
horizon is not exactly the right notion, we prefer the notion of \textit{limited horizon}, which means that the horizon of the agents is not bounded by a number, but that however
it cannot go to infinity.  In the frame of binary games, this means that we forbid paths that go always to the left, we want to consider the paths that eventually go to the right so
that one can compute the utility of each agent.

Since the games are now infinite, a \emph{total function} that associates, by computation, a utility to a strategy can no more be defined, one can only define a \emph{relation}
between a strategy and a utility.  If the path of the choice goes always to the left (on the backbone) the utility cannot be defined, but if the path goes \emph{``eventually to the
  right''}, the utility of a strategy makes sense.  We therefore define a predicate \coqdocid{EvtRight} which says whether the path of the given strategy goes eventually to the
right.  The basic case for this predicate is that it holds for the game \coqdocid{iNode} \coqdocid{a} \coqdocid{right} \coqdocid{sl} \coqdocid{sr}.  The induction case says that if
\coqdocid{EvtRight} \coqdocid{sl} then \coqdocid{EvtRight} (\coqdocid{iNode} \coqdocid{a} \coqdocid{left} \coqdocid{sl} \coqdocid{sr}).  The predicate \coqdocid{EvtRight} is
typical of an inductive predicate.  We get two lemmas which show the existence and the uniqueness of a utility for a strategy that goes eventually to the right, therefore in that
case, the association of the utility to the strategy is functional. Since most of the time we want this functional association, we require the predicate ``eventually right'' to
hold.

\medskip
\noindent
\coqdockw{Lemma} \coqdocid{Existence\_i2u}:  \ensuremath{\forall} (\coqdocid{a}:\coqdocid{Agent}) (\coqdocid{s}:\coqdocid{InfStrategy}),\coqdoceol
\coqdocindent{1.00em}
\coqdocid{EvtRight} \coqdocid{s} \ensuremath{\rightarrow} \ensuremath{\exists} \coqdocid{u}:\coqdocid{Utility}, \coqdocid{i2u} \coqdocid{a} \coqdocid{u} \coqdocid{s}.\coqdoceol

\medskip
\noindent
\coqdockw{Lemma} \coqdocid{Uniqueness\_i2u}: \ensuremath{\forall} (\coqdocid{a}:\coqdocid{Agent}) (\coqdocid{u} \coqdocid{v}:\coqdocid{Utility}) (\coqdocid{s}:\coqdocid{InfStrategy}),\coqdoceol
\coqdocindent{1.00em}
\coqdocid{EvtRight} \coqdocid{s} \ensuremath{\rightarrow} \coqdocid{i2u} \coqdocid{a} \coqdocid{u} \coqdocid{s} \ensuremath{\rightarrow} \coqdocid{i2u} \coqdocid{a} \coqdocid{v} \coqdocid{s} \ensuremath{\rightarrow} \coqdocid{u}=\coqdocid{v}.\coqdoceol
\coqdocindent{0.50em}
\coqdoceol

\medskip

Going eventually to the right is not enough.  In some cases, when dealing with subgames, more specifically with subgame perfect equilibria, one wants to be able to compute utilities in
subgames.  Therefore one wants to be sure that even further in subgames one will go eventually to the right.  Hence we define another predicate that ensures that we will always go
eventually to the right.  This predicate \coqdocid{AlwEvtRight} reminds us the same kind of predicate defined in the frame of temporal logic (see \cite{BertotCasterant04}
chap.13 and~\cite{coupet-grimal03:_axiom_of_linear_temp_logic}).  Since \coqdocid{AlwEvtRight} has to traverse the whole infinite game, it has to be coinductive.  It says that
\coqdocid{AlwEvtRight} (\coqdocid{iNode} \coqdocid{a} \coqdocid{c} \coqdocid{sl} \coqdocid{sr}) holds, if \coqdocid{AlwEvtRight} \coqdocid{sl} and \coqdocid{EvtRight}  \coqdocid{sr} hold, in other
words, a strategy goes always eventually to the right if  its left substrategy goes always eventually to the right and if its right substrategy (which is finite) goes eventually to the right.

\subsection{Convertibility of infinite strategies}
\label{sec:infconv}

The definition of the convertibility \coqdocid{s} \iconva\coqdocid{s'} of two strategies \coqdocid{s} and \coqdocid{s'} is inductive. Its base case is the reflexivity
of the convertibility.  In other words, \coqdocid{s} \iconva\coqdocid{s'} if:
\begin{itemize}
\item \coqdocid{s} = \coqdocid{s'}, or
\item if \coqdocid{s1} \iconva\coqdocid{s1'} and \coqdocid{s2}\conva\coqdocid{s2'} and \coqdocid{s} is \coqdocid{iNode} \coqdocid{a} \coqdocid{c}
  \coqdocid{s1} \coqdocid{s2} and \coqdocid{s'} is \coqdocid{iNode} \coqdocid{a} \coqdocid{c'} \coqdocid{s1'} \coqdocid{s2'} (\coqdocid{c} may be different, but  \coqdocid{a} has to be the same), or
\item if \coqdocid{s1} \iconva\coqdocid{s1'} and \coqdocid{s2}\conva\coqdocid{s2'} and \coqdocid{s} is \coqdocid{iNode} \coqdocid{a'} \coqdocid{c}
  \coqdocid{s1} \coqdocid{s2} and \coqdocid{s'} is \coqdocid{iNode} \coqdocid{a'} \coqdocid{c} \coqdocid{s1'} \coqdocid{s2'} (different agents, but same \coqdocid{c}).
\end{itemize}

\subsection{Equilibria}
\label{sec:eq}

To define Nash Equilibria on infinite games, one must be able to compare utilities.  For that, one must be able to compute those utilities.  Hence one restricts to strategies
that go eventually to the right.

\medskip

\noindent
\coqdockw{Inductive} \coqdocid{InfNashEq}: \coqdocid{InfStrategy} \ensuremath{\rightarrow} \coqdocid{Prop} := \coqdoceol
\noindent
| \coqdocid{INE} : \ensuremath{\forall} (\coqdocid{s}: \coqdocid{InfStrategy}), \coqdoceol
\coqdocindent{1.00em}
\coqdocid{EvtRight} \coqdocid{s} \ensuremath{\rightarrow} \coqdoceol
\coqdocindent{2.00em}
(\ensuremath{\forall} (\coqdocid{a}:\coqdocid{Agent}) (\coqdocid{s'}:\coqdocid{InfStrategy}) (\coqdocid{u} \coqdocid{u'}: \coqdocid{Utility}),\coqdoceol
\coqdocindent{2.50em}
\coqdocid{EvtRight} \coqdocid{s'} \ensuremath{\rightarrow} \coqdocid{s'}\iconva\coqdocid{s} \ensuremath{\rightarrow} (\coqdocid{i2u} \coqdocid{a} \coqdocid{u'} \coqdocid{s'}) \ensuremath{\rightarrow} (\coqdocid{i2u} \coqdocid{a} \coqdocid{u} \coqdocid{s}) \ensuremath{\rightarrow} (\coqdocid{u'} \leut \coqdocid{u})) \ensuremath{\rightarrow}\coqdoceol
\coqdocindent{1.00em}
\coqdocid{InfNashEq} \coqdocid{s}.\coqdoceol

\medskip

On infinite strategies one defines a predicate \emph{Subgame Perfect Equilibrium}.  Because one has to be able to compute utilities on subgames, the Subgame Perfect Equilibrium
predicate, written \coqdocid{SGPE}, is defined on strategies which go always eventually to the right.  Notice that Subgame Perfect Equilibria are defined on a smaller class of
strategies than Nash Equilibria, but also that the definition of \coqdocid{SGPE} is coinductive, since its definition requires to traverse the whole infinite subgame.

\medskip

\noindent
\coqdockw{CoInductive} \coqdocid{SGPE}: \coqdocid{InfStrategy} \ensuremath{\rightarrow} \coqdocid{Prop} :=\coqdoceol
\noindent
| \coqdocid{SGPEnode\_left}: \ensuremath{\forall} (\coqdocid{a}:\coqdocid{Agent})(\coqdocid{u}:\coqdocid{Utility}) (\coqdocid{sl}: \coqdocid{InfStrategy}) (\coqdocid{sr}: \coqdocid{FinStrategy}), \coqdoceol
\coqdocindent{2.00em}
\coqdocid{AlwEvtRight} \coqdocid{sl} \ensuremath{\rightarrow} \coqdocid{SGPE} \coqdocid{sl} \ensuremath{\rightarrow} \coqdocid{BI} \coqdocid{sr} \ensuremath{\rightarrow} \coqdocid{i2u} \coqdocid{a} \coqdocid{u} \coqdocid{sl} \ensuremath{\rightarrow} (\coqdocid{f2u} \coqdocid{sr} \coqdocid{a} \leut \coqdocid{u}) \ensuremath{\rightarrow} \coqdoceol
\coqdocindent{2.00em}
\coqdocid{SGPE} (\coqdocid{iNode} \coqdocid{a} \coqdocid{left} \coqdocid{sl} \coqdocid{sr})\coqdoceol
\noindent
| \coqdocid{SGPEnode\_right}: \ensuremath{\forall} (\coqdocid{a}:\coqdocid{Agent}) (\coqdocid{u}:\coqdocid{Utility}) (\coqdocid{sl}: \coqdocid{InfStrategy}) (\coqdocid{sr}: \coqdocid{FinStrategy}), \coqdoceol
\coqdocindent{2.00em}
\coqdocid{AlwEvtRight} \coqdocid{sl} \ensuremath{\rightarrow} \coqdocid{SGPE} \coqdocid{sl} \ensuremath{\rightarrow} \coqdocid{BI} \coqdocid{sr} \ensuremath{\rightarrow} \coqdocid{i2u} \coqdocid{a} \coqdocid{u} \coqdocid{sl} \ensuremath{\rightarrow} (\coqdocid{u} \leut \coqdocid{f2u} \coqdocid{sr} \coqdocid{a}) \ensuremath{\rightarrow} \coqdoceol
\coqdocindent{2.00em}
\coqdocid{SGPE} (\coqdocid{iNode} \coqdocid{a} \coqdocid{right} \coqdocid{sl} \coqdocid{sr}). \coqdoceol

\medskip

One can show that \coqdocid{InfNashEq} is a fixed point of this definition.  To prove the lemma \coqdocid{SGPE} \coqdocid{s} \ensuremath{\rightarrow} \coqdocid{InfNashEq}
\coqdocid{s}, one needs to perform a proof by induction, this cannot be an induction on the definition \coqdocid{SGPE} which is coinductive.  To enable a proof by induction, we impose an additional
requirement, namely that somewhere in the game a maximal utility is reached for all agents.   We give a statement of this requirement through the following predicate \coqdocid{EvtMaxU}:

\medskip
\noindent
\coqdockw{Inductive} \coqdocid{EvtMaxU}: \coqdocid{InfStrategy} \ensuremath{\rightarrow} \coqdocid{Prop} :=\coqdoceol
\noindent
| \coqdocid{EUM\_basis}: \ensuremath{\forall} (\coqdocid{s}:\coqdocid{InfStrategy}), \coqdoceol
\coqdocindent{1.00em}
(\ensuremath{\forall} (\coqdocid{a}:\coqdocid{Agent})(\coqdocid{u} \coqdocid{u'}:\coqdocid{Utility})(\coqdocid{s'}:\coqdocid{InfStrategy}), \coqdocid{s}\iconva\coqdocid{s'} \ensuremath{\rightarrow} \coqdocid{i2u} \coqdocid{a} \coqdocid{u'} \coqdocid{s'} \ensuremath{\rightarrow} \coqdocid{i2u} \coqdocid{a} \coqdocid{u} \coqdocid{s} \ensuremath{\rightarrow} (\coqdocid{u'} \leut \coqdocid{u})) \ensuremath{\rightarrow} \coqdoceol
\coqdocindent{1.00em} \coqdocid{EvtMaxU} \coqdocid{s}\coqdoceol
\noindent
| \coqdocid{EUM\_gen}:  \ensuremath{\forall} (\coqdocid{a}:\coqdocid{Agent})(\coqdocid{c}:\coqdocid{Choice})(\coqdocid{sl}:\coqdocid{InfStrategy})(\coqdocid{sr}:\coqdocid{FinStrategy}), \coqdoceol
\coqdocindent{1.00em}
\coqdocid{EvtMaxU} \coqdocid{sl} \ensuremath{\rightarrow} \coqdocid{EvtMaxU} (\coqdocid{iNode} \coqdocid{a} \coqdocid{c} \coqdocid{sl} \coqdocid{sr}).\coqdoceol

\medskip

It has two parts.
\begin{itemize}
\item The game fulfills the maximal utility requirement for every agent,

\item The left subgame does.
\end{itemize}

Then we can set and prove the main theorem: 

\medskip
\noindent
\coqdockw{Theorem} \coqdocid{SGPE\_is\_InfNashEq} : \ensuremath{\forall} \coqdocid{s}:\coqdocid{InfStrategy}, \coqdocid{EvtMaxU} \coqdocid{s} \ensuremath{\rightarrow} \coqdocid{SGPE}  \coqdocid{s} \ensuremath{\rightarrow} \coqdocid{InfNashEq} \coqdocid{s}.\coqdoceol

\medskip

If every agent reaches a maximal utility somewhere in the game, then a subgame perfect equilibrium is a Nash equilibrium.

\section{The ``Illogic of Conflict Escalation'' revisited}
\label{sec:escalation}

Now we may want to apply those general results on a specific infinite subgame.  Consider the following game proposed by Shubik~\cite{Shubik:1971}.  Recall its principle. Two agents
\emph{Alice} and \emph{Bob} compete in an auction for an object of a value, say $2$\,\textcent, in this statement.  The two agents bid $2$\,\textcent, one after the other.  If one agent gives up, the
highest bidder gets the object, but the second bidder pays also for his bid.  As Shubik noted this game may never cease.

Let us take as the utility ordering, the order $\ge$ on the \emph{nat} (the natural numbers or non negative integers), in other words, if \emph{u} and \emph{v} are two utilities,
\emph{u} and \emph{v} are of type \coqdocid{nat} and \emph{u} \leut \emph{v} means $u\ge v$, i.e., the larger the bid, the smaller the utility.

\subsection{Strategy \emph{Never give up}}
\label{sec:ngu}

Let us consider the function \coqdocid{enlarge} \coqdocid{left} \coqdocid{left} \coqdocid{n} on strategies, that takes a strategy and returns another one and that can be described by the following picture, where the thick arrows
give the choice at each node.

\bigskip

\(
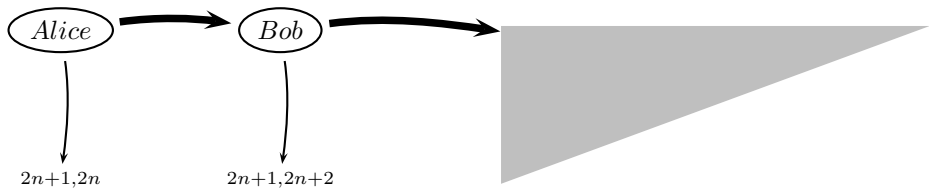
\begin{psmatrix}[nodesep=3pt]
  &{\ovalnode{a}{Alice}}
  &{\ovalnode{b}{Bob}}
  &[name=c] 
  &\\
  &[name=e] {\scriptstyle 2n+1, 2n}
  &[name=f] {\scriptstyle 2n+1, 2n+2}
  &[name=g] \phantom{\scriptstyle 2n+1, 2n+2}
    \ncarc[arrows=->,linewidth=.1]{a}{b}
    \ncarc[arrows=->,linewidth=.1]{b}{c}
    \ncarc[arrows=->]{a}{e}
    \ncarc[arrows=->]{b}{f}
    \pspolygon*[linecolor=lightgray](-.7,0)(-.7,2.1)(5,2.1)
  \end{psmatrix}
\)

\bigskip

We define the strategy \emph{never give up} (in short \emph{ngu}) as the strategy where the agents always keep bidding.  
It requires to  define first a strategy that starts with the value \emph{n} as the solution of 

\begin{center}
  \emph{ngu n = enlarge left left (ngu (S n)).}
  \end{center}

\emph{ngu n} is an infinite strategy and we can prove that for \emph{n}, \emph{(ngu n)} is not a Nash equilibrium.  In this proof, reasoning has a flavor of 
\textit{temporal logic}.

\subsection{Strategy \emph{Always give up}}
\label{sec:agu}

Here we consider the function \coqdocid{enlarge} \coqdocid{right} \coqdocid{right} \coqdocid{n} on strategies.

\bigskip

\(
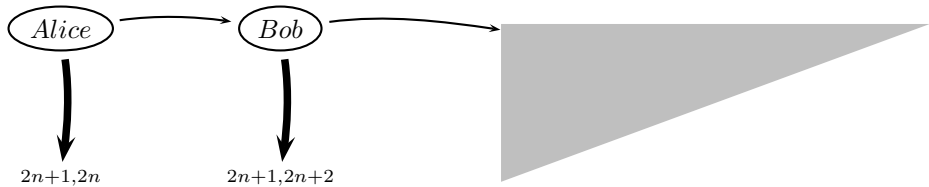
\begin{psmatrix}[nodesep=3pt]
  &{\ovalnode{a}{Alice}} &{\ovalnode{b}{Bob}} &[name=c] &\\
  &[name=e] {\scriptstyle 2n+1, 2n} &[name=f] {\scriptstyle 2n+1, 2n+2} &[name=g] \phantom{\scriptstyle 2n+1, 2n+2} 
  \ncarc[arrows=->]{a}{b} \ncarc[arrows=->]{b}{c}
  \ncarc[arrows=->,linewidth=.1]{a}{e} \ncarc[arrows=->,linewidth=.1]{b}{f} 
  \pspolygon*[linecolor=lightgray](-.7,0)(-.7,2.1)(5,2.1)
\end{psmatrix}
\)

\bigskip

Again \emph{agu n} is an infinite strategy and we can prove the lemma: 

\medskip
\noindent
\coqdockw{Lemma} \coqdocid{SGPEAGU}:  \ensuremath{\forall} (\coqdocid{n}:\coqdocid{nat}), \coqdocid{SGPE} \coqdocid{Ag12} \coqdocid{nat} \coqdocid{ge} (\coqdocid{agu} \coqdocid{n}).\coqdoceol

\medskip
\noindent which says that that this strategy is a \emph{Subgame Perfect Equilibrium.}

\section{Conclusion}
\label{sec:conc}

The experiment presented in this paper is in a very stage.  Its goal is to make clear that, in games, especially in infinite games, the reasoning can be mechanized despite it is
not obvious; it evidences with no surprise some subtlety.  To mimic it, humans elaborate rather complex deductions.  One of the main statement we can make is that true
\emph{classical logic is never used}.  More specifically, we noticed no use of the excluded middle or proofs by double negations.  We hope this experiment opens a discussion and
shows how far humans are from the somewhat ideal mechanized reasoning. We wish to pursue this research by trying other modelings and concepts and going further in the proofs, for
instance finding other conditions for \coqdocid{SGPE} \coqdocid{s} \ensuremath{\rightarrow} \coqdocid{InfNashEq} \coqdocid{s} and deeper results on the example of ``Illogic Conflict of
Escalation''.


\end{document}